\documentclass[aps,prl,twocolumn,superscriptaddress]{revtex4}
\pdfoutput=1

\begin{document}

\pagestyle{empty} 

\noindent {\bf Giomi and Mahadevan Reply:} Starostin and van der Heijden \cite{Starostin:2011} question the validity of the results reported in our Letter \cite{Giomi:2010} regarding the existence of spontaneous helical structure in finite-temperature developable ribbons. According to the authors of the Comment, our setting is affected by a systematic error in the calculation of the torsion $\tau$ that results in a bias of the Monte Carlo sampling toward helical configurations, and originates from the fact that the Frenet frame, used by us to describe the centerline of the ribbons, rotates by 180$^{\circ}$ about the direction of the tangent vector every time an inflection point (i.e. a point where the curvature $\kappa$ changes in sign) is encountered.  

While the rotation of the frame at infection points is a well known property of Frenet frame, the conclusions of Starostin and van der Heijden on how this phenomenon would affect the Monte Carlo sampling are incorrect, as are many of the other statements that they make. To understand why, we start with the free energy density used in our work, given by the Sadowsky functional:
\begin{equation}
f \propto \kappa^{2}+2\tau^{2}+\frac{\tau^{4}}{\kappa^{2}}	
\end{equation}
where $\kappa$ and $\tau$ are respectively the curvature and the torsion of the centerline. We see immediately that configurations with $\kappa=0$ and $\tau\ne 0$ have infinite energy density and therefore zero Boltzmann weight at finite temperature resulting in a suppression of configurations with inflection points, unless they are planar, in which case $\tau = 0$. 

Indeed, inflection points enter in the statistical average only through a single configuration $\kappa=\tau=0$. The latter, however, corresponds to exactly one single point in the entire two-dimensional plane of configurations of  $(\kappa,\tau)$, and therefore occupy a set of zero  Lebesgue measure. Thus inflections do not affect the ensamble integral and consequently the statistics of our simulation. This latter concept can be clarified with an example taken directly from the Comment. Starostin and van der Heijden argue that by experimenting with a strip of paper is not hard to produce an inflected $S$-shaped planar configuration. But the authors fail to realize that the probability of starting from a random conformation of the strip and performing random hand movements to obtain a planar $S$-shaped configuration is zero. This is particularly true in the statistical mechanical setting where the ends of the ribbon are free so that such inflected shapes will never arise. This is because  three-dimensional configurations lead to many more states and are thus entropically more favored: this naturally introduces torsion and thus prevents the formation of any inflections points. Thus one naturally obtains helical configurations as we have shown. The authors further claim that to prevent inflection points, the discretized ribbon must have a torsion that diverges in the continuum limit - this is untrue for the simple reason that there is a finite energy penalty associated with the discrete analogs of curvature and torsion. 

The recalculation of the persistence length, given without much detail in the Comment, is also incorrect. The authors claim to replace the Frenet frame used in our calculation with a more suitable frame, but in fact this is not a properly constructed material frame. Indeed a natural frame is provided by a triad of orthonormal vectors $\{{\bf t},{\bf m}_{1},{\bf m}_{2}\}(s)$ in which ${\bf t}(s)$ is the usual tangent vector and ${\bf m}_{1}(s)$ and ${\bf m}_{2}(s)$ are constructed at any point $s$ along the curve by parallel transport of an arbitrarily chosen pair of orthonormal vector at the origin: ${\bf m}_{1}(0)$ and ${\bf m}_{2}(0)$ as discussed for example in Ref. \cite{Rabin:2000}.  Additionally, they still make use of the Frenet-Serret equations in their discrete form ${\bf F}_{i}={\bf F}_{i-1}{\bf R}_{i}$ to describe the motion of their frame along the centerline and calculate the various thermal averages, an inconsistency that is at odds with their claim to not use the standard Frenet frame.

For the reasons given above, the criticisms of Starostin and van der Heijden Comment are without basis. Our results that free developable ribbons in three dimensions at finite temperature posses an underlying helical structure remain valid.
\\[15pt]
L. Giomi and L. Mahadevan\\
{\small
School of Engineering and Applied Sciences\\
Harvard University\\
Cambridge, Massachusetts 02138, USA}

\end{document}